%% file: DCIS.tex
\documentclass{imammb}

\jno{dqnxxx}

\usepackage{natbib}
\usepackage{graphicx}
\usepackage{amsmath,amsthm,bm,mathrsfs}
\usepackage{amssymb}

\input standard.tex

\numberwithin{equation}{section}
\newcommand{\beq}{\begin{equation}}
\newcommand{\eeq}{\end{equation}}

\newcommand{\R}{\mathbb R}

\newcommand{\deltadist}{{\delta}}
\newcommand{\defeq}{{\stackrel{\mbox{\tiny{def}}}{=}}}

\begin{document}

\title{The time-evolution of DCIS size distributions with applications to breast cancer growth and progression}
\author{
{\sc James G. Dowty}\\[2pt]
University of Melbourne, \\
Level 3 / 207 Bouverie St, Carlton, 3053, Australia.\\[6pt]
{\sc Graham B. Byrnes}\\[2pt]
International Agency for Research on Cancer, \\
150 Cours Albert Thomas, 69372 Lyon cedex 08, France.\\[6pt]
{\sc Dorota M. Gertig}\\[2pt]
Victorian Cytology Service, \\
752 Swanston Street, Carlton, 3053, Australia.\\[6pt]
{\rm [Received on 23 May 2012]}\vspace*{6pt}}
\pagestyle{headings}
\markboth{DOWTY, BYRNES \& GERTIG}{\rm THE TIME-EVOLUTION OF DCIS SIZE DISTRIBUTIONS}
\maketitle

\begin{abstract}
{Ductal carcinoma {\em in situ} (DCIS) lesions are non-invasive tumours of the breast which are thought to precede most invasive breast cancers (IBC).  As individual DCIS lesions are initiated, grow and invade (i.e. become IBC) the size distribution of the DCIS lesions present in a given human population will evolve.  We derive a differential equation governing this evolution and show, for given assumptions about growth and invasion, that there is a unique distribution which does not vary with time.  Further, we show that any initial distribution converges to this stationary distribution exponentially quickly.  It is therefore reasonable to assume that the stationary distribution is equal to the true DCIS size distribution, at least for human populations which are relatively stable with respect to the determinants of breast cancer.  Based on this assumption and the size data of 110 DCIS lesions detected in a mammographic screening program between 1993 and 2000, we produce maximum likelihood estimates for certain growth and invasion parameters.  Assuming that DCIS size is proportional to a positive power $p$ of the time since tumour initiation we estimate $p$ to be $0.50$ with a $95\%$ confidence interval of $(0.35, 0.71)$.  Therefore we estimate that DCIS lesions follow a square-root growth law and hence that they grow rapidly when small and relatively slowly when large.  Our approach and results should be useful for other mathematical studies of cancer, especially those investigating biological mechanisms of invasion.}
{Breast cancer, cancer progression, DCIS, population dynamics, maximum likelihood}
\end{abstract}

\section{Introduction}

Most breast cancers arise in the epithelial cells which line
the milk ducts of each breast (\cite{DCISsurvey,breast}). While a
tumour remains confined to the system of ducts it is known as a
{\em ductal carcinoma in situ} (DCIS) of the breast.  If the
tumour penetrates the epithelial membrane which bounds the ductal
system then it is termed an {\em invasive breast cancer} (IBC).
In this paper we will use the term {\em tumour}
to mean either DCIS or IBC and we use {\em DCIS lesion} or simply {\em lesion} as a synonym for DCIS.
IBCs are considered to be life-threatening whereas DCIS lesions {\em per se} are not, since the  epithelial membrane
prevents DCIS lesions from spreading to critical organs.

The transition from DCIS to IBC is known as {\em invasion}.  It is not known what biological events trigger invasion or what proportion of DCIS
lesions would invade if left untreated because essentially all DCIS lesions are surgically removed once discovered.
However, as this surgery can be psychologically and physically damaging and could potentially spread cancerous cells, determining the risk
associated with untreated DCIS is of great clinical importance.

Currently it is not ethically possible to directly observe the natural history of DCIS
(though see \cite{editorial}) so our knowledge comes from indirect sources: follow-up
studies of DCIS initially misdiagnosed as benign; studies of
recurrence of DCIS as invasive cancer; autopsy studies; studies of
risk factors for DCIS; animal studies; and mathematical studies
(\cite{DCISsurvey}).  As \cite{DCISsurvey} point out, all of these studies
have weaknesses for estimating the proportion of DCIS lesions which invade, e.g. lesions
observed in follow-up studies are likely to be of low histological grade and consequently
less prone to invasion than other DCIS lesions.  In this context, a mathematical study of DCIS, despite its
reliance on simplifying assumptions, may be able to make an important
contribution to the ongoing debate (\cite{editorial}) about the clinical significance
of DCIS.

Previous mathematical studies of breast cancer have generally
focussed on IBC.  Some of these studies do not incorporate any DCIS phase (\cite{atkinson,hart})
while others include a DCIS phase only as a period of latency between tumour initiation and the
start of invasive growth (\cite{kopans,moolgavkar,yakovlev}).  By contrast, the present paper
is chiefly concerned with DCIS.  \cite{franks} also explicitly study DCIS
though their approach is very different to the one presented here.

DCIS {\em size} is defined as the maximal linear extent of the lesion, or its diameter.  While
volume is often used to describe the size of IBCs, the linear or branched shape of DCIS makes this
definition impractical (\cite{breast}).

The rest of this paper is set out as follows.  In \S \ref{S:stable} we use the results of \cite{diekmann} and the example of
\cite{hart} to give an equation governing the time-evolution of the DCIS size distribution
of a given human population as individual lesions grow and invade.  For given assumptions about growth and invasion there is a unique {\em stationary distribution}
which does not vary with time.  Further, we prove (in \S \ref{S:stability}) that any initial
DCIS size distribution converges to this stationary distribution exponentially quickly.  Therefore,
for a human population which is relatively stable with respect to the determinants of breast cancer,
it is reasonable to assume that the stationary distribution is close to
the true DCIS size distribution.  We use this assumption in \S \ref{S:lik} and \S \ref{S:estimation}
to generate maximum likelihood estimates for certain DCIS growth and invasion parameters, based on
data from a mammographic screening program. In \S \ref{S:conclusion} we draw some conclusions and
finish with some ideas for future research.

\section{The stationary distribution of DCIS size}
\label{S:stable}

In this section we find an equation governing the evolution of the size distribution of DCIS lesions present in a human population as individual lesions are initiated, grow and invade.  For given growth and invasion rates there is a unique {\em stationary distribution}
which does not vary with time and we give an expression for this distribution.  As mentioned above, this section is based on the results of \cite{diekmann} and motivated by the ideas of \cite{hart}.

\cite{diekmann} derive an equation governing the time-evolution of the size distribution of a population of cells which grow, die and reproduce by binary fission.  Setting the fission rate to zero gives an equation describing the size distribution of a population of individuals which grow and stochastically leave the population, so this applies to a population of DCIS lesions which grow and invade.  So following \cite{diekmann} (though using slightly different notation) we assume that individual DCIS lesions of size $x$ grow and invade at rates $f(x)/\alpha$ and $\mu(x)$, respectively.  Hence the size $y(\tau)$ of a DCIS lesion at time $\tau$ evolves according to the differential equation
\begin{eqnarray}
\label{E:ODE} \frac{dy}{d\tau}=  \frac{1}{\alpha}f(y)
\end{eqnarray}
and if $X$ is a random variable giving the tumour size at invasion (conditional on
$\alpha$) then $Pr(X>x)$ is
\begin{equation}\label{E:Gintegral}
    G(x) = \exp \left( -\int_0^x \frac{\mu(\xi) \alpha}{f(\xi)} \,d\xi \right)
\end{equation}
({\em cf.} (2.4) of \cite{diekmann}).

The function $f$ determines the DCIS growth law and we assume it is such that any solution $y : [0,\infty) \to \R$ of (\ref{E:ODE}) with $y(0)$ approximately the size of one epithelial cell is a strictly monotonically increasing function.  The {\em growth factor} $\alpha > 0$ is constant for each tumour.  For the rest of this section we treat this parameter as a constant but in later sections we will allow it to vary randomly from tumour to tumour.

Now, let $\phi(x,t)$ be an (unnormalized) density function for the DCIS lesions in the population, i.e. suppose that $\phi$ is such that $\int_0^x \phi(\xi,t)\, d\xi$ is the expected number of lesions at time $t$ with size in the range $[0,x]$, for any $x>0$ and $t$.  Then applying equation (2.1) of \cite{diekmann} to our situation (where the fission rate $b$ is zero) gives
\begin{eqnarray}
\label{E:balancelaw1}
\frac{\partial \phi}{\partial t} + \frac{\partial ~}{\partial x} \left(\frac{\phi f}{\alpha} \right)
= -\mu \phi.
\end{eqnarray}

So far we have not accounted for the process of tumour initiation.  However a straight-forward modification of the argument in the Appendix of \cite{diekmann} shows that if tumours of size $x$ are initiated at a rate $S(x)$ then $\phi$ will evolve according to
\begin{eqnarray}
\label{E:balancelaw2}
\frac{\partial \phi}{\partial t} + \frac{\partial ~}{\partial x} \left(\frac{\phi f}{\alpha} \right)
= -\mu \phi + S.
\end{eqnarray}
The source term $S$ is unknown but since DCIS lesions are
initially approximately the size of one epithelial cell we know that
$S(x) = 0$ for all $x > \epsilon$ where $\epsilon > 0$ is microscopically small.

Using (\ref{E:Gintegral}) to write $\mu$ in terms of $G$ we arrive at the following
inhomogenous linear equation governing the time-evolution of $\phi$:
\begin{eqnarray}
\label{E:dphidt}
\frac{\partial \phi}{\partial t} + \frac{f}{\alpha} \frac{\partial \phi}{\partial x} +
\left(\frac{f^\prime}{\alpha} - \frac{fG^\prime}{\alpha G}\right) \phi = S.
\end{eqnarray}

Now, let $\psi$ be the {\em unnormalized stationary distribution}, meaning that it satisfies (\ref{E:dphidt}) and is a function of $x$ only, i.e.
${\partial \psi}/{\partial t} = 0$.  Then for all macroscopic $x$,
(\ref{E:dphidt}) becomes
\begin{eqnarray*}
 0 =  \psi \frac{f}{\alpha} \left(
 - \frac{\psi^\prime}{\psi} - \frac{f^\prime}{f} +\frac{G^\prime}{G} \right)
\end{eqnarray*}
where the primes denote differentiation with respect to $x$.  This has the unique solution
\begin{eqnarray}
\label{E:phi} \psi(x) = c \,\frac{G(x)}{f(x)}
\end{eqnarray}
where $c$ is a constant.  If $c$ is chosen so that $\int_0^\infty \psi(x) \, dx = 1$ then we refer to $\psi$ as the (normalized) {\em stationary distribution}.

\section{Convergence to the stationary distribution}
\label{S:stability}

In \S \ref{S:stable} we used the results of \cite{diekmann} to show that, for given growth and invasion rates, there is a unique DCIS size distribution
which does not vary with time.  Unfortunately the convergence results of \cite{diekmann} cannot be directly applied to our situation of continuous tumour initiation and zero fission rates ({\em cf.} the assumptions $H_b$ on page 229 of \cite{diekmann}).  Therefore, in this section we prove that any distribution converges to the stationary distribution exponentially quickly (and with a reasonable time-scale).  This result adds legitimacy to our use of the
stationary distribution in \S \ref{S:estimation} to produce maximum likelihood estimates of DCIS growth and invasion parameters.  However the proof is fairly technical and a reader who is prepared to accept this result can skip this section without
compromising his or her understanding of the later sections.

We now fix a value of $\alpha$ for the rest of this section and we use the notation of \S \ref{S:stable} throughout except that we write $\phi_x$ and
$\phi_t$ for the partial derivatives of $\phi$.  In this notation, (\ref{E:dphidt}) becomes
\begin{eqnarray}
\label{E:inhomogPDE}  \phi_t +  \left(\frac{f}{\alpha}\right) \phi_x +
\left(\frac{f^\prime}{\alpha} - \frac{fG^\prime}{\alpha
G}\right) \phi = S.
\end{eqnarray}
We will show that for any function $\phi$ which satisfies
(\ref{E:inhomogPDE}), the sequence of functions $\phi(\cdot,t)$ converges to the unnormalized stationary distribution $\psi$ as $t \to \infty$.  The proof will be given for the model of invasion used in \S \ref{S:lik} and \S \ref{S:estimation}, where $\mu(x)= h$ is constant, but the proof extends to other models without major modifications.

It follows from (\ref{E:inhomogPDE}) that $\deltadist \defeq \phi - \psi$ satisfies the homogenous partial
differential equation
\begin{eqnarray}
\label{E:homogPDE} \deltadist_t + \left(\frac{f}{\alpha}\right) \deltadist_x +
\left(\frac{f^\prime}{\alpha} - \frac{fG^\prime}{\alpha
G}\right) \deltadist = 0.
\end{eqnarray}
Our aim is to show that $\delta(\cdot,t)$ approaches $0$ as $t \to \infty$.  Since (\ref{E:homogPDE}) is a linear, first-order partial differential equation we can solve this equation by the method of
characteristics (e.g. see \cite{john} pages 8--14). The graph
$$ \{ (x,t,z) \in \R^3 \mid  z = \deltadist(x,t) \}$$
of a solution $\deltadist$ of (\ref{E:homogPDE}) is known as an {\em
integral surface}, and the images of the integral curves of the
vector field
\begin{eqnarray}
\label{E:VF}  \left(\frac{f(x)}{\alpha},1, \frac{z
f(x)G^\prime(x)}{\alpha G(x)} - \frac{z
f^\prime(x)}{\alpha}\right)
\end{eqnarray}
are the {\em characteristic curves} of (\ref{E:homogPDE}).  The
projection of a characteristic curve into the $xt$-plane is called
a {\em characteristic}. Our interest in the characteristic curves
derives from the fact that every integral surface is a union of
characteristic curves (\cite{john}).

Now, let $\deltadist_0: \R_{>0} \to \R$ be given, where $\R_{>0}$
denotes the positive reals.  From now on we let $\deltadist$ denote
the solution of (\ref{E:homogPDE}) which satisfies the initial
conditions $\deltadist(\xi,0) = \deltadist_0(\xi)$ for every $\xi > 0$.  Let $(x,t) \in \R_{>0} \times \R$ be given and assume for the rest of this section that $y$ is the solution of (\ref{E:ODE}) with $y(t) = x$.  The characteristic
which passes through the point $(x,t)$ is the graph of $y$, i.e. the characteristic is
the trajectory of a tumour which is of size $x$ at time $t$. If
the characteristic does {\em not} meet the positive $x$-axis then
the initial conditions above do not determine $\deltadist(x,t)$. In this case we set $\deltadist(x,t)=0$ since no tumour in existence at time $0$ has size $x$ at time $t$ and there is no source term in
(\ref{E:homogPDE}).  On the other hand, if the characteristic
through $(x,t)$ meets the positive $x$-axis then the initial conditions determine $\deltadist(x,t)$ as follows.

It is easily checked that any integral curve $\gamma$ of (\ref{E:VF}) is given by
\begin{eqnarray}
\label{E:intcurve} \gamma(\tau) =  \left(Y(\tau+t_0),\tau+t_0,z_0
\frac{f(x_0) G(Y(\tau+t_0))}{G(x_0) f(Y(\tau+t_0))}\right)
\end{eqnarray}
for some $x_0,t_0,z_0 \in \R$ where the function $Y$
solves (\ref{E:ODE}) with $Y(t_0) = x_0$. However, since the
corresponding characteristic curve (i.e. the image of $\gamma$) does not depend on
$t_0$ we may assume without loss of generality that $t_0=0$.  Suppose now that the image of $\gamma$ is the characteristic curve which passes through $(x,t,\deltadist(x,t))$.  This implies firstly that $Y(t) = x$ and hence $Y=y$ and secondly that $\gamma$ lies in the integral surface of $\deltadist$, so from $\gamma(0) = (x_0,0,z_0)$ we have $z_0 = \deltadist_0(x_0)$.  Combining this with (\ref{E:intcurve}) gives
\begin{eqnarray}
\label{E:deltaxt1} \deltadist(x,t)=\deltadist_0(I(x,t)) \frac{
G(x)}{G(I(x,t)) } \frac{f(I(x,t)) }{f(x)}
\end{eqnarray}
where we have written $I(x,t)$ in the place of $x_0$ to make its dependence on $x$ and $t$ explicit. Here $I(x,t)$ is the initial size of a tumour which is of size $x$ at time $t$, i.e. $I(x,t)$ is the function of $x$ and $t$ which is determined by
the condition $I(x,t) = y(0)$ where, as above, $y$ solves (\ref{E:ODE}) with $y(t) = x$.

As in \S \ref{S:lik} we assume that the time between tumor initiation and invasion is a random variable $T$ with constant hazard $h$, i.e. that $Pr(T > t) = \exp(-ht)$.  Let $X$ be the tumour size at invasion and recall that $G(\xi) = Pr(X > \xi)$.  Then for a tumour which was initiated at time $a$, $X = y(T+a)$ and so
$$G(\xi) = Pr(X > \xi) = Pr(y(T+a) > \xi) = Pr(T>y^{-1}(\xi) - a) = \exp(-h(y^{-1}(\xi) - a)).$$
This is consistent with (\ref{E:Galpha}) because $a=0$ in \S \ref{S:lik}.  Therefore, regardless of the value of $a$,
\begin{eqnarray}
\label{E:GxGx0} \frac{G(x)}{G(I(x,t))} = e^{-ht}.
\end{eqnarray}

Now, from (\ref{E:ODE}) we have
$$  t/\alpha = \int_0^t \frac{d\tau}{\alpha} = \int_{I(x,t)}^x \frac{d\xi}{f(\xi)} = F(x) - F(I(x,t))$$
where $F$ is any function which satisfies $F^\prime = 1/f$.  This equation holds for arbitrary $x$ and $t$ so taking partial derivatives with respect to $x$ and rearranging gives
\begin{eqnarray}
\label{E:dx0dx} \frac{f(I(x,t))}{f(x)} = \frac{\partial I}{\partial x}(x,t).
\end{eqnarray}
So by combining (\ref{E:deltaxt1}), (\ref{E:GxGx0}) and (\ref{E:dx0dx}) we obtain
\begin{eqnarray}
\label{E:convergence}
\int_0^\infty | \deltadist(\xi,t) | d\xi =  e^{-ht} \int_0^\infty |
\deltadist_0(\iota) | d\iota.
\end{eqnarray}
This shows that the functions $\deltadist(\cdot,t)$ approach $0$ exponentially quickly with respect to the $L^1$ norm as $t \to \infty$.  Therefore any DCIS size distribution approaches the stationary distribution exponentially quickly as $t \to \infty$.

Since $h$ is the hazard of invasion, the time-scale of convergence
to the stationary distribution is comparable to the time for an individual DCIS lesion to invade, perhaps of the order of 20 years.  Hence the rate of convergence is comparable to the rate of  population changes which affect the incidence of breast cancer such as changes to the distributions of:  age at first birth; age at menarche; pre- and post-menopausal body-mass index; and exogenous hormone use.  So in human populations where the determinants of breast cancer have changed slowly in recent years, we would expect the population DCIS size distribution to track with the changes in the underlying stationary distribution without ever deviating too far from it.

The above arguments are conditional on $\alpha$, however since the rate of convergence in (\ref{E:convergence})
does not depend on $\alpha$, the same convergence results apply when we marginalize over $\alpha$ as in the next section.

\section{Parametric assumptions}
\label{S:lik}

In \S \ref{S:estimation} we will use the method of maximum likelihood to estimate certain important features of DCIS growth and invasion.  As for most likelihood-based estimation procedures we must first make some parametric assumptions.  In this section we describe these assumptions and give an explicit expression for the corresponding stationary distribution.

We assume first that DCIS size is proportional to a positive power of the time since tumour initiation.  This corresponds to the choice
\begin{eqnarray}
\label{E:power}f(y) = ky^\beta
\end{eqnarray}
in (\ref{E:ODE}) where $k >0$ and $\beta < 1$ are constants.  Then
(\ref{E:ODE}) has the solution
\begin{eqnarray}
\label{E:powergrowth}y(t) = [kt(1-\beta)/\alpha +
x_0^{1-\beta}]^{\frac{1}{1-\beta}}
\end{eqnarray}
where $x_0$ is the initial size of the DCIS lesion (see equation
(2) of \cite{hart}).  Therefore the inverse of $y$ is given by $y^{-1}(x) = [x^{1-\beta} -
x_0^{1-\beta}] \alpha/(k(1-\beta))$.  Assuming that $x_0$ is negligibly small compared to $x$ (which is reasonable, since $x_0$ is approximately the size of one epithelial cell) we have
\begin{eqnarray}
\label{E:yinv}y^{-1}(x) =  \frac{\alpha x^{1-\beta}}{k(1-\beta)}.
\end{eqnarray}

Now, as in \S \ref{S:stable}, let $X$ be a random variable representing tumour size at
invasion conditional on the growth factor $\alpha$ and define
$G(x) = Pr(X > x)$.  Let $T$ be the time from tumour initiation to invasion
conditional on $\alpha$, i.e. $T = y^{-1}(X)$ where $y^{-1}$ is as given in (\ref{E:yinv}). We
model invasion by the simple assumption that the hazard function
for $T$ is some constant $h > 0$, i.e. that
$$Pr(t \le T < t + \Delta t \mid t \le T) =
h \Delta t + O(\Delta t^2),$$ so $Pr(T > t) = \exp(-ht)$.
Combining this with $G(x) = Pr(T > y^{-1}(x))$ gives
\begin{eqnarray}
\label{E:Galpha} G(x) = \exp\left[-h y^{-1}(x)\right].
\end{eqnarray}
Substituting (\ref{E:yinv}) into (\ref{E:Galpha}) then
substituting the result into (\ref{E:phi}) gives an expression for the stationary distribution $\psi_\alpha$ conditional on $\alpha$:
\begin{eqnarray}
\label{E:phi1} \psi_\alpha(x) = c_\alpha x^{-\beta}
\exp\left[\frac{-h \alpha x^{1-\beta}}{k(1-\beta)}\right]
\end{eqnarray}
where $c_\alpha$ does not depend on $x$.   The integral
$\int_0^\infty \psi_\alpha(x) dx$ exists for $\beta < 1$ and
substituting (\ref{E:phi1}) into the equation $\int_0^\infty
\psi_\alpha(x) dx = 1$ and effecting a simple change of variables
shows
$$c_\alpha = \alpha h/k.$$

Now, as in \cite{brown} we take the growth factor $\alpha$
to be exponentially distributed. Due to the presence of $k$ in (\ref{E:power})
we may assume without loss of generality that the expected value of
$\alpha$ is $1$. Marginalising (\ref{E:phi1}) over $\alpha$
gives the PDF of the (unconditional) stationary distribution:
\begin{eqnarray}
 \psi(x) &=& \int_0^\infty e^{-\alpha} \psi_\alpha(x) d\alpha \nonumber \\
&=& x^{-\beta} h k^{-1}  \int_0^\infty \alpha \exp\left[-
\alpha \left(\frac{h x^{1-\beta}}{k(1-\beta)} + 1\right)\right] d\alpha \nonumber \\
&=& x^{-\beta} k h^{-1} \left[ k h^{-1} +
x^{1-\beta}(1-\beta)^{-1} \right]^{-2} \label{E:phi_explicit}
\end{eqnarray}
where the last step uses integration by parts.

\section{Comparison with mammography data}
\label{S:estimation}

In \S \ref{S:stability} we showed that any DCIS size distribution approaches the stationary
distribution exponentially fast (and on a relatively short time-scale).  Therefore it seems reasonable to assume that the size distribution
of DCIS lesions present in a real human population will approximate the stationary distribution corresponding to the processes of growth and invasion in operation.
In this section we use this assumption, together with the parametric assumptions of \S \ref{S:lik}, to obtain maximum likelihood
estimates for the growth and invasion parameters introduced in \S \ref{S:lik}.  This estimation will be based on DCIS size data
collected by a mammographic screening program operated by BreastScreen Victoria (BSV), in Victoria, Australia since 1993.

\subsection{The BSV dataset}

The BreastScreen Victoria (BSV) data was collected between 1993
and 2000. In the Australian state of Victoria, women are invited to
attend free breast screen clinics every 2 years between the ages
of 50 and 69 years. We restricted our data to this age range and
to DCIS lesions which were detected via a woman's first mammogram.
The ages at first mammogram were approximately uniformly
distributed across this age range.  The size of a DCIS lesion was
measured by a pathologist after it was detected via a mammogram
and surgically removed. Most DCIS lesions are roughly linear in
shape, and for such lesions, DCIS size is essentially the length
of the lesion.

\begin{figure}[!htbp]
\begin{center}
\includegraphics[height=8cm]{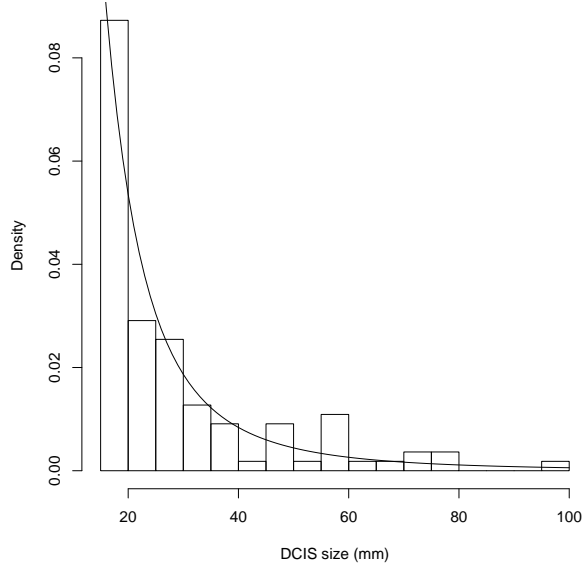}
\end{center}
\caption{A histogram of DCIS size in the BSV dataset, with the
fitted DCIS size distribution overlaid} \label{F:hist}
\end{figure}

Since very few pre-invasive breast cancers are detected outside of
mammographic screening programs, the BSV dataset is an excellent
source of data on larger DCIS lesions. However, detection of a
small DCIS lesion via mammographic screening is uncertain, and
probably depends on a number of factors such as the
orientation of the lesion, its position in a dense or non-dense
part of the breast and the presence or absence of
micro-calcification (\cite{breast}). Therefore, in order to avoid these complications, we restricted the
BSV data to those lesions with size greater than or equal to 15mm
(though see \S \ref{Sub:sens}).  We also excluded two outlying
DCIS lesions with size greater than 100mm. This gave us a
dataset of $110$ DCIS lesions of size at least 15mm which were
detected on a woman's first mammogram when she was between $50$
and $69$ (inclusive) years old.  A histogram of these sizes is
given in Figure \ref{F:hist}.

\subsection{Maximum likelihood estimates}
\label{Sub:MLests}

We would now like to use the PDF (\ref{E:phi_explicit}) of the stationary distribution
 to obtain maximum likelihood estimates.
However since we have restricted the BSV data to tumours with size
greater than or equal to 15mm we must first condition this PDF
appropriately.  A simple change of variables shows
\begin{eqnarray*}
 \int_{15}^\infty \psi(x) dx
&=& \int_{15}^\infty x^{-\beta} k h^{-1}  \left[ k h^{-1} +
x^{1-\beta}(1-\beta)^{-1} \right]^{-2} dx \\
&=&  \left[  1 + \frac{ 15^{1-\beta} }{k h^{-1} (1-\beta)}
\right]^{-1}
\end{eqnarray*}
so the appropriate PDF for the BSV data is
\begin{eqnarray}
\label{E:cond_PDF}
 \left[  1 + \frac{ 15^{1-\beta} }{k h^{-1}(1-\beta)}
 \right]  \frac{x^{-\beta}k h^{-1} } {\left[ k h^{-1} +
x^{1-\beta}(1-\beta)^{-1} \right]^{2}}.
\end{eqnarray}
Note that $h$ and $k$ only appear in (\ref{E:cond_PDF}) in the
combination $k h^{-1}$.  Therefore it is not possible to produce
separate maximum likelihood estimates for the parameters $h$ and $k$, only for
the quantity $\theta \defeq k h^{-1}$.  Therefore we define the
following likelihood function
\begin{eqnarray}
\label{E:Lik} L(\beta,\theta) = \prod_{i = 1}^{110}
 x_i^{-\beta} \frac{  [\theta +
 {15}^{1-\beta}(1-\beta)^{-1}] } {[ \theta +
 x_i^{1-\beta}(1-\beta)^{-1}]^{2}}
\end{eqnarray}
where $x_1, x_2, \ldots, x_{110}$ are the sizes (in millimeters)
of the DCIS lesions in the BSV dataset.

We re-parameterized the likelihood (\ref{E:Lik}) in terms of
$\beta$ and $\log \theta$ then calculated maximum likelihood
estimates using the `mle' function of R 2.11.1 (\cite{R}).  The corresponding estimates for $\beta$ and $\theta$ are $$(\hat{\beta}, \hat{\theta}) = (-0.99,30.9).$$
The PDF (\ref{E:cond_PDF})
corresponding to these estimates is overlaid on the histogram of
the BSV data in Figure \ref{F:hist}, showing a good fit.

We generated conservative 95\% confidence intervals (95\%~CI) for $\beta$ and $\theta$ by
projecting the region of the parameter space where the relative
log-likelihood is greater than $-5.99$ onto the $\beta$ and $\theta$
co-ordinate axes (see \S 4.5 of \cite{severini}). This gave a 95\%~CI
of $(-1.70,-0.48)$ for $\beta$ and $(0.00,1278)$ for $\theta$.

\subsection{Sensitivity analysis}
\label{Sub:sens}

We varied the size cutoff, introduced because of the uncertain detection of small lesions, in order to assess its impact on our estimates.  Conditioning on DCIS size greater than or equal to 10mm, 15mm or 20mm gave approximately the same estimates for $\beta$, as can be seen from Table \ref{table1}.

\begin{table}[!b]
\caption{Results of a sensitivity analysis in which the size cutoff is varied}
\begin{tabular}{llll}
\hline\noalign{\smallskip}
Cutoff (mm) & Number of DCIS lesions & $\hat{\beta}$ (95\% CI) & $\hat{\theta}$ (95\% CI) \\
\noalign{\smallskip}\hline\noalign{\smallskip}
  10 & 155 & -1.07 (-1.70, -0.47)  & 86 (5.9, 661)  \\
  15 & 110 & -0.99 (-1.86, -0.41)  & 31 (0.00, 1278)   \\
  20 & 77  & -1.06 (-2.13, -0.55)  & 0.41 (0.00, 4357) \\
\noalign{\smallskip}\hline
\end{tabular}
\label{table1}
\end{table}

\subsection{Interpretation of the results}

The confidence interval for $\theta$ is so wide that we cannot draw any conclusions regarding
the growth or invasion rates of DCIS lesions.  However the estimate for $\beta$ is more reliable.  By (\ref{E:powergrowth}) our estimate $\hat{\beta}=-0.99$ corresponds, to a good approximation,
to DCIS growth in which the lesion size is proportional to the square root of the time since tumour initiation.
The bounds of the 95\%~CI place the exponent between $0.35$ and $0.71$, corresponding to a class of curves with similar, very rapid initial
growth which then slows with increasing time and DCIS size.

This may be biologically plausible
for two reasons: (1) unlike large lesions, small lesions are
unconstricted by the epithelial membrane and surrounding tissues which bound the
milk-ducts; and (2) it may be that large lesions receive less
nutrients per cell than small ones due to limited contact with the epithelial membrane (\cite{franks}).

Even if the assumption of a power-law for growth is overly restrictive, our results still suggest that DCIS growth slows down as the lesion increases in size, in contrast to IBC which grows at an increasing rate (\cite{hart}).

\begin{figure}[!htbp]
\begin{center}
\includegraphics[height=8cm]{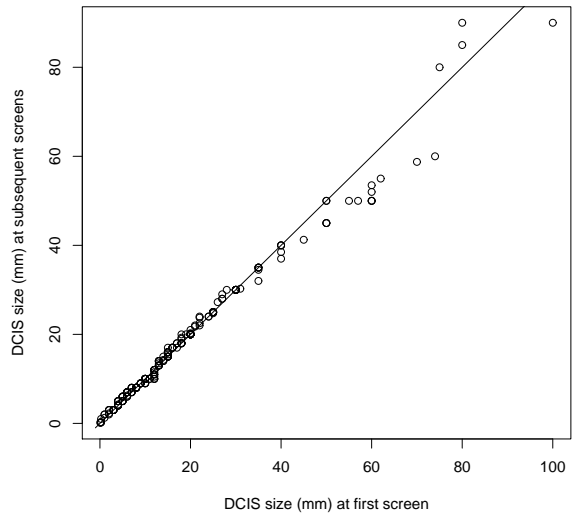}
\end{center}
\caption{A quantile-quantile plot comparing the DCIS size distributions of lesions which are detected on a woman's first mammographic screening with those which are detected on a subsequent screen.  }
\label{F:qqplot}
\end{figure}

A square-root growth function may also be the key to explaining a
surprising feature of the BSV data.  The size
distribution for DCIS lesions detected on a woman's first
mammographic screening is remarkably similar to the size
distribution for tumours detected on a subsequent screen (see Figure \ref{F:qqplot}).
This is surprising because first-screen-detected tumours could have been growing for more than 20 years while
subsequent-screen-detected tumours can only have been present and
of a detectable size for 2 years (the usual time between
screens).  Our estimate that DCIS growth is initially
very rapid and then relatively slow may help to explain this feature.
Note that the estimates of \S \ref{Sub:MLests} were only based on the size data of DCIS lesions detected on the first screen.

\section{Conclusions and future research}
\label{S:conclusion}

Because the natural history of DCIS cannot ethically be observed,
indirect studies of DCIS growth and invasion such as the present
one are needed to resolve the controversies surrounding the
clinical significance of DCIS.  The main contribution of our paper to this debate is our estimate that
DCIS lesions approximately obey a square-root growth function and that
they grow rapidly when small and relatively slowly when large.
However, we have also shown how the unobserved processes of DCIS growth and invasion affect the
observed DCIS size distribution, and we hope that this approach will also be useful for future studies of DCIS.

One limitation of our approach is that, due to the form of the likelihood function, it is
impossible to produce separate estimates for the rates of DCIS
growth and invasion (see \S \ref{Sub:MLests}).  However, it may be possible to overcome this limitation by
incorporating information on DCIS lesions which are detected at a woman's second or later mammographic screening.
Without such an extension, even with a larger dataset, it is probably not possible to use our approach to estimate
the proportion $P$ of DCIS lesions which will invade within a
lifetime.  This proportion is of great clinical importance because if it is very small then the current treatment regime
of mandatory removal may be causing more harm than good, as suggested recently~(\cite{editorial}).
In any case, we hope that our estimate of the DCIS growth law will be used in
other mathematical studies of DCIS and that one of these will
produce credible estimates for $P$.

Another possible use of our approach is as a test on
competing theories of invasion.  At present, the
biological events which trigger invasion are not known, however different biological invasion mechanisms will generally give different forms for the
distribution of DCIS size at invasion. For example, if it is assumed that invasion
occurs when a DCIS lesion grows along a milk duct to a
pre-existing `hot spot' (which might be a weakness in the
epithelial membrane or a mutant stromal cell) then rather than (\ref{E:Galpha}) we would
have $G(x) = \exp(-hx)$ for some constant $h>0$.  It may therefore be possible
to use the methods of this paper to test different biological
theories of invasion by computing maximum likelihood estimates and comparing the parsimony of the different models using a measure such as the Bayesian information criterion.

\end{document}

%% file: standard.tex
 \newtheoremstyle{theorem}{6pt}{6pt}{\rm}{}{\sffamily}{ }{ }{}
 \theoremstyle{theorem}

 \newtheoremstyle{algorithm}{6pt}{6pt}{\rm}{}{\sffamily}{ }{ }{}
 \theoremstyle{algorithm}

 \newtheoremstyle{lemma}{6pt}{6pt}{\rm}{}{\sffamily}{ }{ }{}
 \theoremstyle{lemma}

\newtheoremstyle{case}{6pt}{6pt}{\rm}{}{\sffamily}{. }{ }{}
 \theoremstyle{case}

 \newtheoremstyle{statement}{6pt}{6pt}{\rm}{}{\sffamily}{ }{ }{}
\theoremstyle{statement}

 \newtheoremstyle{corollary}{6pt}{6pt}{\rm}{}{\sffamily}{ }{ }{}
 \theoremstyle{corollary}

  \newtheoremstyle{definition}{6pt}{6pt}{\rm}{}{\sffamily}{ }{ }{}
 \theoremstyle{definition}

\newtheoremstyle{example}{6pt}{6pt}{\rm}{}{\sffamily}{ }{ }{}
\theoremstyle{example}

\newtheoremstyle{remark}{6pt}{6pt}{\rm}{}{\sffamily}{ }{ }{}
\theoremstyle{remark}

\newtheoremstyle{approximation}{6pt}{6pt}{\rm}{}{\sffamily}{ }{ }{}
\theoremstyle{approximation}

\newtheoremstyle{scheme}{6pt}{6pt}{\rm}{}{\sffamily}{ }{ }{}
\theoremstyle{scheme}

\newtheoremstyle{Algorithm}{6pt}{6pt}{\rm}{}{\sffamily}{ }{ }{}
\theoremstyle{Algorithm}

\newtheoremstyle{Assumption}{6pt}{6pt}{\rm}{}{\sffamily}{ }{ }{}
\theoremstyle{Assumption}

\newtheoremstyle{proposition}{6pt}{6pt}{\rm}{}{\sffamily}{ }{ }{}
\theoremstyle{proposition}

\newtheoremstyle{hypo}{6pt}{6pt}{\rm}{}{\sffamily}{ }{ }{}
 \theoremstyle{hypo}

  \newtheoremstyle{Step}{6pt}{6pt}{\rm}{}{}{ }{ }{}
 \theoremstyle{Step}